\documentclass[12pt]{JHEP}
\usepackage{psfrag}
\usepackage{graphics}

\newcounter{multieqs}

\newcommand{\bq}{\begin{equation}}
\newcommand{\fq}{\end{equation}}
\newcommand{\bqr}{\begin{eqnarray}}
\newcommand{\fqr}{\end{eqnarray}}

\newcommand{\be}{\begin{equation}}
\newcommand{\ee}{\end{equation}}
\newcommand{\eq}[1]{(\ref{#1})}

\newcommand{\bra}[1]{\langle #1|}
\newcommand{\ket}[1]{|#1 \rangle}

\newcommand{\bm}[1]{\mbox{\boldmath $#1$}}

\def\bd{\begin{document}}
\def\ed{\end{document}}
\def\nn{\nonumber}
\def\bea{\begin{eqnarray}}
\def\eea{\end{eqnarray}}
\let\bm=\bibitem
\let\la=\label
\renewcommand{\theequation}{\thesection.\arabic{equation}}
\def\npb#1#2#3{Nucl. Phys. {\bf{B#1}} #3 (#2)}
\def\plb#1#2#3{Phys. Lett. {\bf{#1B}} #3 (#2)}
\def\prl#1#2#3{Phys. Rev. Lett. {\bf{#1}} #3 (#2)}
\def\prd#1#2#3{Phys. Rev. {D \bf{#1}} #3 (#2)}
\def\cmp#1#2#3{Comm. Math. Phys. {\bf{#1}} #3 (#2)}
\def\cqg#1#2#3{Class. Quantum Grav. {\bf{#1}} #3 (#2)}
\def\nppsa#1#2#3{Nucl. Phys. B (Proc. Suppl.) {\bf{#1A}}#3 (#2)}
\def\ap#1#2#3{Ann. of Phys. {\bf{#1}} #3 (#2)}
\def\ijmp#1#2#3{Int. J. Mod. Phys. {\bf{A#1}} #3 (#2)}
\def\rmp#1#2#3{Rev. Mod. Phys. {\bf{#1}} #3 (#2)}
\def\mpla#1#2#3{Mod. Phys. Lett. {\bf A#1} #3 (#2)}
\def\jhep#1#2#3{J. High Energy Phys. {\bf #1} #3 (#2)}
\def\atmp#1#2#3{Adv. Theor. Math. Phys. {\bf #1} #3 (#2)}

\newcommand{\EQ}[1]{\begin{equation} #1 \end{equation}}
\newcommand{\AL}[1]{\begin{subequations}\begin{align} #1 \end{align}\end{subequations}}
\newcommand{\SP}[1]{\begin{equation}\begin{split} #1 \end{split}\end{equation}}
\newcommand{\ALAT}[2]{\begin{subequations}\begin{alignat}{#1} #2 \end{alignat}\end{subequations}}
\def\beqa{\begin{eqnarray}}
\def\eeqa{\end{eqnarray}}
\def\beq{\begin{equation}}
\def\eeq{\end{equation}}

\def\N{{\cal N}}
\def\sst{\scriptscriptstyle}
\def\thetabar{\bar\theta}
\def\Tr{{\rm Tr}}
\def\one{\mbox{1 \kern-.59em {\rm l}}}

\def\a{\alpha}          \def\da{{\dot\alpha}}
\def\b{\beta}           \def\db{{\dot\beta}}
\def\c{\gamma}  \def\C{\Gamma}  \def\cdt{\dot\gamma}
\def\d{\delta}  \def\D{\Delta}  \def\ddt{\dot\delta}
\def\e{\epsilon}                \def\vare{\varepsilon}
\def\f{\phi}    \def\F{\Phi}    \def\vvf{\f} \def\vphi{\varphi}
\def\h{\eta}
\def\k{\kappa}
\def\l{\lambda} \def\L{\Lambda}
\def\m{\mu}     \def\n{\nu}
\def\o{\omega}
\def\p{\pi}     \def\P{\Pi}
\def\r{\rho}
\def\s{\sigma}  \def\S{\Sigma}
\def\t{\tau}
\def\th{\theta} \def\Th{\Theta} \def\vth{\vartheta}
\def\X{\Xeta}
\def\z{\zeta}

\def\cA{{\cal A}} \def\cB{{\cal B}} \def\cC{{\cal C}}
\def\cD{{\cal D}} \def\cE{{\cal E}} \def\cF{{\cal F}}
\def\cG{{\cal G}} \def\cH{{\cal H}} \def\cI{{\cal I}}
\def\cJ{{\cal J}} \def\cK{{\cal K}} \def\cL{{\cal L}}
\def\cM{{\cal M}} \def\cN{{\cal N}} \def\cO{{\cal O}}
\def\cP{{\cal P}} \def\cQ{{\cal Q}} \def\cR{{\cal R}}
\def\cS{{\cal S}} \def\cT{{\cal T}} \def\cU{{\cal U}}
\def\cV{{\cal V}} \def\cW{{\cal W}} \def\cX{{\cal X}}
\def\cY{{\cal Y}} \def\cZ{{\cal Z}}

\def\ua{\underline{\alpha}}
\def\ub{\underline{\phantom{\alpha}}\!\!\!\beta}
\def\uc{\underline{\phantom{\alpha}}\!\!\!\gamma}
\def\um{\underline{\mu}}
\def\ud{\underline\delta}
\def\ue{\underline\epsilon}
\def\una{\underline a}\def\unA{\underline A}
\def\unb{\underline b}\def\unB{\underline B}
\def\unc{\underline c}\def\unC{\underline C}
\def\und{\underline d}\def\unD{\underline D}
\def\une{\underline e}\def\unE{\underline E}
\def\unf{\underline{\phantom{e}}\!\!\!\! f}\def\unF{\underline F}
\def\unm{\underline m}\def\unM{\underline M}
\def\unn{\underline n}\def\unN{\underline N}
\def\unp{\underline{\phantom{a}}\!\!\! p}\def\unP{\underline P}
\def\unq{\underline{\phantom{a}}\!\!\! q}
\def\unQ{\underline{\phantom{A}}\!\!\!\! Q}
\def\unH{\underline{H}}

\def\As {{A \hspace{-6.4pt} \slash}\;}
\def\bs {{b \hspace{-6.4pt} \slash}\;}
\def\Ds {{D \hspace{-6.4pt} \slash}\;}
\def\ds {{\del \hspace{-6.4pt} \slash}\;}
\def\ss {{\s \hspace{-6.4pt} \slash}\;}
\def\ks {{ k \hspace{-6.4pt} \slash}\;}
\def\ps {{p \hspace{-6.4pt} \slash}\;}
\def\pas {{{p_1} \hspace{-6.4pt} \slash}\;}
\def\pbs {{{p_2} \hspace{-6.4pt} \slash}\;}

\def\Fh{\hat{F}}
\def\Vh{\hat{V}}
\def\Xh{\hat{X}}
\def\ah{\hat{a}}
\def\xh{\hat{x}}
\def\yh{\hat{y}}
\def\ph{\hat{p}}
\def\xih{\hat{\xi}}

\def\psit{\tilde{\psi}}
\def\Psit{\tilde{\Psi}}
\def\tht{\tilde{\th}}

\def\At{\tilde{A}}
\def\Qt{\tilde{Q}}
\def\Rt{\tilde{R}}

\def\ft{\tilde{f}}
\def\st{\tilde{s}}
\def\pt{\tilde{p}}
\def\qt{\tilde{q}}
\def\vt{\tilde{v}}

\def\delb{\bar{\partial}}
\def\bz{\bar{z}}
\def\Db{\bar{D}}

\def\bk{{\bf k}}
\def\bl{{\bf l}}
\def\bp{{\bf p}}
\def\bq{{\bf q}}
\def\br{{\bf r}}
\def\bx{{\bf x}}
\def\by{{\bf y}}
\def\bR{{\bf R}}
\def\bV{{\bf V}}

\def\d{\delta}\def\D{\Delta}\def\ddt{\dot\delta}

\def\pa{\partial} \def\del{\partial}
\def\xx{\times}

\def\trp{^{\top}}
\def\inv{^{-1}}
\def\dag{{^{\dagger}}}
\def\pr{^{\prime}}

\def\rar{\rightarrow}
\def\lar{\leftarrow}
\def\lrar{\leftrightarrow}

\newcommand{\0}{\,\!}      
\def\one{1\!\!1\,\,}
\def\im{\imath}
\def\jm{\jmath}

\newcommand{\tr}{\mbox{tr}}
\newcommand{\slsh}[1]{/ \!\!\!\! #1}

\def\vac{|0\rangle}
\def\lvac{\langle 0|}

\def\hlf{\frac{1}{2}}
\def\ove#1{\frac{1}{#1}}

\def\Box{\square}
\def\ZZ{\mathbb{Z}}
\def\CC#1{({\bf #1})}
\def\bcomment#1{}
\def\bfhat#1{{\bf \hat{#1}}}
\def\VEV#1{\left\langle #1\right\rangle}

\newcommand{\ex}[1]{{\rm e}^{#1}} \def\ii{{\rm i}}

\def\vs{\vspace*{.3cm}}

\title{Hermitian Analyticity, IR/UV Mixing 
and Unitarity of Noncommutative Field Theories}

\author{Chong-Sun Chu$^{a}$, Jerzy Lukierski$^{b}$
\footnote{Supported by KBN grant 5P03B05620}
and Wojtek J. Zakrzewski$^{a}$\\
$^a$Centre for Particle Theory, Department of Mathematical Sciences,
University of Durham, Durham, DH1 3LE, UK\\
$^b$Institute for Theoretical Physics,
University of Wroc{l}aw,
pl. M. Borna 9, 50-205 Wroc{l}aw, Poland }

\abstract{
The IR/UV mixing and the violation  of unitarity 
are two of the most intriguing aspects of noncommutative quantum field theories.
In this paper the relation between these two phenomena is
explained and established 
in an explicit form.
We start out by showing that the $S$-matrix of noncommutative field theories is
hermitian analytic. As a consequence, a
noncommutative field theory is unitary if the discontinuities 
of its Feynman diagram amplitudes agree
with the expressions calculated using the Cutkosky formulae. 
These unitarity constraints relate the discontinuities of amplitudes
with physical intermediate states; and allow us to see how the IR/UV mixing
may lead to a breakdown of unitarity.
Specifically, we show 
that the IR/UV singularity does not lead to the violation of unitarity in the
space-space noncommutative case, but it does lead to its
violation in a space-time noncommutative field theory.
As a corollary, noncommutative field theory without 
IR/UV mixing will
be unitary in both the space-space and space-time
noncommutative case.
To illustrate this, we introduce and analyse
the noncommutative Lee model--an exactly solvable quantum field theory. 
We show that the model is free from the IR/UV mixing in both the space-space
and space-time noncommutative cases. Our
analysis is exact. 
Due to absence of the IR/UV mixing one can expect that the theory is
unitary. We present some 
checks supporting this claim.  
Our analysis provides a counter example to the generally held
beliefs that field theories with space-time noncommutativity are
non-unitary.
}

\keywords{Non-Commutative Geometry, Unitarity, Analyticity, S-Matrix } 
\preprint{{\tt hep-th/0201144}}    
\begin{document}  

\section{Introduction}

Recently there has been a lot of activities in constructing and understanding
field theories on noncommutative spacetime (see e.g.
 \cite{doug1,szab2}). There are many reasons why such approaches are of
interest, most of them   related to the desire to take into consideration the
 quantum gravity effects and to understand the nature of spacetime
at very short distances (see e.g. \cite{dopl3,gara4}).
Some of the most recently considered
noncommutative geometries are the noncommutative Minkowski space
 $\bR^{D-1,1}$ \cite{CDS,DH,CH1,sch,sw}, 
the fuzzy sphere  $S_N^2$ \cite{mado6,grosse95,cms}, 
and the $\k$-Minkowski spacetime  \cite{zak8,maj9,luk10}.
The algebra of functions on noncommutative $\bR^{D-1,1}$ is generated by
noncommutative space--time coordinates $\hat{x}^\m$ obeying the
commutation relations ($\mu,\nu = 0,1, ... D-1$).
\be \label{cr1} 
[\hat{x}^\m,\hat{x}^\n] = i \th^{\m\n},
\ee
where $\th^{\m\n}$ is an anti-symmetric constant matrix. The fuzzy
sphere  $S_N^2$ is generated by Hermitian operators
$\hat{x}= ( \hat{x}_1, \hat{x}_2, \hat{x}_3)$
satisfying the defining relations ($i,j,k=1,2,3$).
\be [ \hat{x}_i, \hat{x}_j ] = i\l_N  \e_{ijk} \,\hat{x}_k, 
\qquad \hat{x}_1^2 + \hat{x}_2^2 +\hat{x}_3^2  = R^2.
\ee
Here the
noncommutativity parameter $\l_N $ has the dimension of length  and
should be taken positive. The radius $R$ of the fuzzy sphere is
quantized, in units of $\l_N$, by
\be\label{def3} \frac {R}{\l_N} =
\sqrt{\frac{N}{2} \left( \frac{N}{2} +1\right) } \; , \quad
\mbox{$N = 1,2,\cdots$ } \ee
The $\k$-Minkowski spacetime is
defined by the basic relations between the three commuting space
coordinates ( $[ \hat{x}_i, \hat{x}_j]=0$)
and a noncommutative quantum 
time variable $\widehat{t}\, (\widehat{x}_0 = c\widehat{t} )$:
\be 
[\hat{x}_0 ,\hat{x}_i] = \frac{i}{\k} \hat{x}_i.
\ee

In this paper we consider the case of noncommutative
$\bR^{D-1,1}$.  This topic has been studied extensively
(for a recent review, see e.g. \cite{doug1,szab2}  and references therein).
Field theory on this noncommutative space can be  obtained  by  the
replacement of standard products of  fields by the Moyal $*$-product
induced by the relation  (\ref{cr1})
\footnote{
We denote $x=( {\bf x} , t)$,
$ z' =  ( {\bf z'},\tau' )$, $z'' = ( {\bf z''} , \tau'' )$ and 
use the notation $a_\mu \theta^{\mu\nu} b_\nu \equiv a\theta b$.
},
\be \label{luchu1.5}
A\cdot B(x) \longrightarrow A \ast B(x)
= e^{-i \frac{\th}{2}  \frac{\del}{\del z'} \frac{\del}{\del z''}}
A(x + z') B (  x + z'') |_{z' = z'' = 0} .
\ee
In the momentum basis, the result of such an operation is 
the appearance of an additional Moyal phase factor $V(k^1, \cdots,k^N)$
\be\label{moyal}
 e^{ik^1 x}* e^{ik^2 x} * \cdots  *  e^{ik^N x} = V(k^1, \cdots,k^N) \;
e^{i\sum_i k^i x},\quad 
V(k^1, \cdots,k^N):= 
e^{\sum_{i\leq j} \frac{i}{2}k^i \th k^j}.
\ee
Due to  this phase factor one has to fix a definite cyclic
ordering (say, anti-clockwise) 
of the momenta that enter any vertex of a given Feynman diagram.

An intriguing phenomenon for the quantum field theory on  
noncommutative $\bR^{D-1,1}$ is the existence of an infrared/ultraviolet
(IR/UV) mixing \cite{uvir} in the quantum effective
action. Due to this
mixing, IR singularities arise from integrating out the UV degrees of
freedom. This threatens the renormalizability and even the
consistency  of a QFT on noncommutative $\bR^{D-1,1}$.
Hence a better understanding (beyond the technical level) of 
the mechanism of IR/UV mixing and possible ways to resolve it are certainly highly
desirable. 
We recall that so far in the literature, field theory on
noncommutative $\bR^{D-1,1}$ has been quantized  by following the standard
perturbative procedures: namely, the action is expanded around the free
action and  the  corresponding Feynman rules are then written down. This
 is justified in the commutative case; however, since the introduction
of $\th^{\m\n}$ necessarily breaks the Lorentz symmetry from $SO(D-1,1)$
to a smaller group that is left unbroken by the  
commutation relations \eq{cr1}, it is actually
quite unnatural to employ  the standard perturbative vacuum,
i.e. the one defined by the free action and so respecting
the 
{\it full} Lorentz symmetry.
This leads one to suspect that the IR/UV mixing may be
reflecting only the properties of the perturbation theory, and may be
altered or disappear completely
in the full nonperturbative regime (see for example, \cite{Griguolo}). 
An exactly solvable field theory would be a good  ground for testing this idea 
\cite{ag}. This leads us to introduce and study  the noncommutative Lee model.

Another intriguing phenomenon for any quantum field theory on 
noncommutative spacetime is that unitarity could be violated.
It is commonly believed that noncommutative field theory with
space-space  noncommutativity is unitary,  
while theory with space-time  noncommutativity is not. This is
consistent with the fact that space-space  noncommutative field theory
can be embedded in string theory \cite{sw,bcr,dorn,kl1,shenker,hong}, 
while field theory with space-time
noncommutativity cannot \cite{sst,gmms,bara}.
In \cite{gomis} it was found  that the unitarity 
constraints (see \eq{img}) are satisfied for
noncommutative theories with space noncommutativity but are violated for
theories with a noncommuting time (see 
also \cite{agbz,mateos,Bassetto,pmho} for  
recent discussions). However these constraints are, in general, 
actually a stronger statement than the 
unitarity itself. 
The constraints presume a symmetric condition (see \eq{symmetric}) 
which is not generally valid. Without making any additional assumptions,
in this paper, we examine directly the analyticity and
unitarity of the $S$-matrix of a general noncommutative field theory.
We  show that Feynman amplitudes of a noncommutative theory are
hermitian analytic (see \eq{HA}), a useful characterization of the $S$-matrix
as introduced and proven by Olive \cite{olive}. As a result, the
statement that the $S$-matrix is unitary takes the boundary-analytic 
form \eq{discT}; and
that the discontinuity of a Feynman diagram amplitude  
can be computed according to the Cutkosky formulae \cite{cutkosky}.  
 
Although these two phenomena have received a lot of attention and have been
throughly discussed in the literature,  as far as we know, 
the relation between them has not been identified 
explicitly and explained before.
One of the main aims of this paper is to identify and explain such a 
relation between IR/UV singularity and the 
possible violation of unitarity in a
noncommutative field theory. This relation will be established through the
boundary-analytic
unitarity constraints \eq{discT}. The basic idea is that the unitarity 
constraints allow one to relate the discontinuity of a scattering
amplitude in a physical region with the appearance of intermediate
states that can be put on-shell in this region. However,
 in a noncommutative theory, 
IR singularities can also be generated due to the IR/UV mixing. These new
singularities do not correspond to any physical intermediate degrees of
freedom. So, generally, one can expect that the unitarity constraints could 
be violated. In this paper we show, that in the case of space-space
noncommutativity, the new IR singularities are safe in the sense that
they do not generate any discontinuities in the scattering amplitudes.
 However, the IR
singularities do generate such discontinuities in the space-time
noncommutative case.  This is the basic field theoretic mechanism for
the violation of unitarity  in a noncommutative theory. 
We stress that this violation of unitarity occurs only  if
time is noncommuting 
{\it and} in the presence of singularities due to the IR/UV mixing.

To illustrate the above ideas, we introduce and analyse the
noncommutative Lee model. 
Lee model  \cite{lee} is an exactly solvable, nonrelativistic model.
The noncommutative Lee model can  be defined by using the deformed
product of fields \eq{luchu1.5}. The model remains exactly solvable.
We  show that the noncommutative Lee model is free from the IR/UV
mixing both at the perturbative level, and in the full exact
answer. Thus the noncommutative Lee model does not provide a
resolution of the IR/UV mixing issue. This may appear to be
disappointing from the
point of view of looking for a nonperturbative resolution of the IR/UV 
mixing issue. 
Nevertheless, the absence of an IR/UV singularity in a noncommutative field
theory is nontrivial. This is one of the main results of this paper.
Moreover, due to the absence of the IR/UV mixing, one can expect, from the
the above mentioned general arguments, that the Lee model with 
space-time noncommutativity is unitary.
We  provide some further arguments to  support this claim.

The plan of our presentation is as follows:
In section 2.1, we review some basic facts about the $S$-matrix of
commutative field theory. In section 2.2,
we  prove that Feynman diagram amplitudes in a
noncommutative field theory are hermitian analytic and we investigate 
the consequences of this statement on the unitarity of the theory. 
We show that the usual form of the unitarity constraints used by 
many people is not correct in general. 
We derive the correct form of the unitarity constraints and show how they
can be used to check  the unitarity of a given noncommutative theory.
In section 2.3, we explain how a IR/UV singularity may
lead to a breakdown of unitarity in  space-time noncommutative field
theory. 
In section 3, we  study the issue of the IR/UV mixing and unitarity in the
noncommutative Lee model. 
In section 3.1 we describe the commutative Lee model. We show that this
model is renormalizable with the renormalization constants easily
computed in a closed form. It is well known that the original Lee model
in 4-dimensional spacetime has a ghost state and is not unitary  
\cite{lee,pauli,ford}. We discuss improved versions of the original Lee
model that do not have these problems; 
and restrict ourselves to these models when we introduce
noncommutativity and address the issue of the unitarity of the noncommutative model.
This we do in section 3.2 where 
we introduce the space-space noncommutative and the space-time  
noncommutative Lee model via the substitutions \eq{luchu1.11} and
\eq{luchu4.1}. We show that there is no IR/UV mixing in either case
and one can expect that the theory is unitary. 
We present some arguments supporting this claim.

\section{Unitarity and Hermitian Analyticity}

In this section, we  discuss some useful properties of the
$S$-matrix. We refer the reader to  \cite{olive2} and to the 
excellent monograph \cite{elop} for further details on this subject.
We follow the notations and nomenclature of \cite{elop}.

\subsection{$S$-Matrix in the Commutative Case}

First we consider the commutative case. Unitarity of a quantum field
theory follows from the existence of a hermitian Hamiltonian. In terms
of the onshell $S$-matrix, unitarity is the statement that 
\be
S S^\dag = S^\dag S =1.
\ee  
Due to the cluster decomposition property of the $S$-matrix, it is
meaningful to decompose $S$ into two parts
\be
S = 1+ i T,
\ee
with $T$ is the transition matrix.
Written in terms of $T_{ab} := \bra{a} T \ket{b}$, we have 
\be \label{uni}
T_{ab} - T_{ba}^* = i \sum_{n} T_{na}^* T_{nb}= i \sum_{n} T_{an} T_{bn}^*,
\ee 
where  the sum  is over all intermediate states
associated with putting particles onshell.
The $S$-matrix and the transition matrix $T$ 
are defined for external particles with real momenta. 
Since both are invariant under proper Lorentz transformations 
their matrix elements (transition amplitudes)
must be functions of Lorentz scalars which can be formed out of
the momenta.
We call a combination of external lines of the amplitude for a given
physical process  a {\it channel}, and two channels whose lines are disjoint
and exhaustive a {\it reaction}. For an amplitude with $n$ external
lines, there are $2^{n-1} -n-1$ different reactions provided that we
exclude reactions with  single-particle channels and do not
distinguish the direction of the reaction.
The channel invariant variable is the square of the energy in the
given channel $C$,
\be
s=s_C = - (\sum_{i \in C} \pm p_i)^2
\ee
where $\pm p_i$ are the momenta of incoming and outgoing lines,
respectively. $s_C$'s are generalizations of the Mandelstam $s,t,u$
variables for $2 \to 2$ scattering. It is convenient to discuss
the singularity structure of a scattering amplitude in terms of the
space of these $2^{n-1} -n-1$ different channel invariants. For more details
see: \cite{olive2}. 

The transition amplitudes typically have singularities.  
In perturbation theory, the transition amplitude $T_{ab}$ 
is given by the sum of a number of Feynman diagrams $M_{ab}$, each
corresponding to a different channel. The
Feynman integral is typically of the form
\be
I_G (p) =  \int \prod_l^L d^D k_l \prod_i^I \frac{i}{q_i^2 -m_i^2} \cdot B ,
\ee
where  $B$ is  a real normalization 
factor that contains the couplings and factors of $\pi$, $i$ etc and
$p$'s are the external momentum.
As we have said before, the integral can be written in terms of the $s$'s.
If one extends $s$ to the complex plane, then the singularities 
are typically branch points in the complex $s$-plane
\footnote{
The locations of the singularities are determined by the Landau equations,
see for example \cite{elop}. 
We remark that the Landau equations are entirely fixed in
terms of the singularity  manifold $\cT$ of the {\it integrand} of the
Feynman integral, and since noncommutativity modifies the
integrand by a phase factor, the Landau equations 
are unmodified by noncommutativity.
}. 
Extending $s$ to the complex domain, 
one can think of $T_{ab}$ (or $M_{ab}$) as the boundary value of an
analytic function defined on the complex $s$-plane. The resulting
analytic function has singularities on the real $s$-axis that
correspond to physically accessible momenta. These singularities are
called the {\it physical region singularities}.
In addition, this analytic
function may have  additional singularities that correspond to
external momenta that are not physically accessible. 
The analysis of these additional singularities is
more complicated and is not usually performed.

The existence of singularities in the amplitude is a consequence of
unitarity \cite{olive}.  The reasoning is that as the channel invariant
increases past a certain threshold (in the physical region of the considered
amplitude) that corresponds to a new possible intermediate
state, a new term enters the unitarity equation and this gives rise
to a singularity in that channel. Such singularities are called
{\it normal thresholds}. 
The physical region is divided into segments by the normal
thresholds singularities.
It can shown, within perturbation theory, that the amplitudes in these segments can be
continued consistently into the complex plane and be related
analytically if one adopts in the Feynman integrals
the $+ i\e$ prescription by replacing $ m\sp2 \to m \sp2- i\e$, $ \e>0$. This corresponds
to associating an $+i \e$ with a channel invariant when it is close to a
normal threshold. The $+ i\e$ prescription in the correct invariant is
appropriate for all physical region normal thresholds in all
amplitudes \cite{elop}. 
Furthermore it can be shown that the Feynman amplitudes (and hence also $T$) are
{\it hermitian analytic} \cite{olive}, i.e. they satisfy:
\be \label{HA}
M_{ab}(s)^* = M_{ba}(s^*).
\ee

As a consequence of the hermitian analyticity \eq{HA}, the
unitarity relation \eq{uni} can be put in a more elegant form
\be \label{discT}
{\rm Disc}\; T_{ab} = i \sum_{n} T_{an}^{(-)} T_{nb}^{(+)} =  
i \sum_{n} T_{an}^{(+)} T_{nb}^{(-)}.
\ee
Here $f^{(\pm)}$ denotes the boundary values, on the real axis, 
respectively from above and below
the cut, of a complex function $f$,
\be
f^{(\pm)}(s) : = \lim_{\e \to 0^+} f(s \pm i\e), 
\quad  s \in \bR,
\ee
and ${\rm Disc}\; f$ is the  discontinuity across this cut
\be
{\rm Disc}\; f := f^{(+)} -f^{(-)} .
\ee
The relation \eq{discT} is actually somewhat stronger. Indeed, as a 
result of unitarity and hermitian analyticity, it holds
for each individual Feynman diagram  \cite{cutkosky}
\be \label{discM}
{\rm Disc}\; M_{ab} = i \sum_{n} M_{an}^{(-)} M_{nb}^{(+)} =  
i \sum_{n} M_{an}^{(+)} M_{nb}^{(-)}.
\ee
In \eq{discT} and \eq{discM} the discontinuities in a given
channel of the amplitude are associated with  normal thresholds.

In terms of Feynman diagrams, the matrix elements $M_{ab}^{(\pm)}$ are
given, respectively, in terms of the $\pm i \e$ prescription: $m\sp2 \to
m\sp2 \mp i \e$. The RHS of \eq{discM} can be computed using the
``cutting rules'' of Cutkosky \cite{cutkosky}: first cut the diagram in all
possible ways such that the cut propagators can go on shell
simultaneously (for a given set of $s$'s), then, for each cut, replace 
the propagators by $-2\pi i \d(p^2 -m^2)$ in the relativistic case, and by
$-2\pi i\d(p_0 -E(\bp,m))$ in the nonrelativistic case. Finally 
sum the contributions of all possible cuts. 

Before we embark on the noncommutative case, let us remark that
the equation \eq{uni} is sometimes written in  the form \cite{chew}
(or for $M$),
\be \label{img}
2 \; {\rm Im} T_{ab}  = \sum_{n} T_{na}^* T_{nb}.
\ee
To arrive at this form, the following symmetric relation
\be \label{symmetric}
T_{ab} =T_{ba}
\ee
has been assumed. This relation holds,
for example, when the theory is $T$-invariant
and rotationally invariant, and the basis vectors
$\ket{a}$ are chosen to be eigenstates of the total angular
momentum \cite{schweber}. However, we
would like to stress that this relation is not true in general. Failure of
\eq{img} can be due to either the symmetry condition
\eq{symmetric} or the unitarity of the theory \eq{uni}  not being 
satisfied
or if the amplitude possesses singularities which are not due
to the possible intermediate states.  Therefore, generically, \eq{img} is not a conclusive check
of whether a given theory is unitary or not. 
In the next subsection we show that the 
hermitian analyticity  remains valid 
in the noncommutative case
and, therefore, that \eq{discT} and \eq{discM} can be used to check unitary of a
noncommutative theory.

\subsection{$S$-Matrix in the Noncommutative Case} 

In a noncommutative quantum field theory the propagators take the same
form  as in the commutative case while the vertices are modified by 
the Moyal phase factor \eq{moyal} that arises from  
the noncommutative multiplication.
For example, in the noncommutative $\phi^3$ model, 
the modification  of the (real) coupling is a multiplication  by a real factor
\be
g \to g \cos (\frac{1}{2} p \th k),
\ee  
where $k$ and $p$ are the momenta entering the vertex.
However, it is easy to see that 
when the theory involves more fields, 
the modification of the vertex is, generally, a phase factor. For
example, this is the case for the noncommutative Lee model to be
introduced in the next section.  The phase
factor \eq{moyal} is cyclically symmetric but not permutation
symmetric. Therefore, the symmetric relation is, in
general, not valid.

Since Lorentz invariance is broken, 
in addition to the channel invariants we have introduced above, the
$S$-matrix of a noncommutative field theory generally depends also on 
the variables
\be \label{st} 
\tilde{s}_C = - (\sum_{i \in C} \pm \tilde{p_i})^2, \quad \tilde{p_i} : = \th p_i.
\ee

A novelty in noncommutative theory is the 
possible existence of the IR/UV mixing
\cite{uvir}, which states that the amplitudes in a
noncommutative theory become singular in the $\st=0$ limit as one removes the
cutoff, i.e. $\L \to \infty$. These singularities occur in the
physical region of momenta but 
do not correspond to normal thresholds 
since the IR/UV singularities are not related to any new degrees of
freedom. One may  extend the amplitude analytically to above the cut
associated with these singularities by adding $+i \e$ to $\st$. This
corresponds to extending the $i \e$ prescription for the Feynman
diagram to the cutoff: $\L^2 \to \L^2 + i \e$ since the combination
$1/\L^2 - \st$ often appears together \cite{uvir}. 

\vs
\underline{Hermitian analyticity}
\vs
 
Next we examine the hermitian analyticity of a noncommutative Feynman diagram. 
We  show that the Feynman amplitudes for
noncommutative theories are hermitian analytic. To see this, we
note that under the complex conjugation, the Moyal phase factor
\eq{moyal} becomes 
\be \label{reverse}
V(k^1,k^2,\cdots,k^N)^* = V(k^N, \cdots,k^2,k^1),
\ee
i.e. it reverses  the cyclic ordering of the momenta entering the
vertex. 
We can interpret the RHS as the Moyal phase factor of a vertex which
is the mirror image of the original one, see figure 1. 
In the operator language the 
RHS of \eq{reverse} corresponds
to a Wick contraction in the reverse  order. For example,
\bea
M_{ab} &\sim \bra{0}a_1 a_2 (\bar{\phi_1}* \bar{\phi_2}* \phi_3) a_3\dag \ket{0}
& \sim V(k_1,k_2,k_3) \nn\\
M_{ba} &\sim \bra{0}a_3 ( \bar{\phi_3}*\phi_2*\phi_1) a_1\dag a_2\dag \ket{0}
& \sim V(k_3,k_2,k_1) =  V(k_1,k_2,k_3)^*,
\eea
where $\ket{a} = a_1\dag(k_1) a_2\dag(k_2) \ket{0}, \ket{b} =
a_3\dag(k_3) \ket{0}$  in this example. In general, let
\be
V^G :=\prod_{v\in G} V_v
\ee
be the product of  the Moyal phase factors associated with the vertices $v$
of a Feynman diagram $G$.
We have 
\be \label{Vstar}
(V^{G})^* = V^{\bar{G}},
\ee
where $\bar{G}$ is the mirror diagram of $G$. 

\FIGURE{
\label{mirror}
\psfrag{k1}{$k_1$}
\psfrag{k2}{$k_2$}
\psfrag{k3}{$k_3$}
\psfrag{kN}{$k_N$}
\psfrag{G}{$G$}
\psfrag{GB}{$\bar{G}$}
\includegraphics{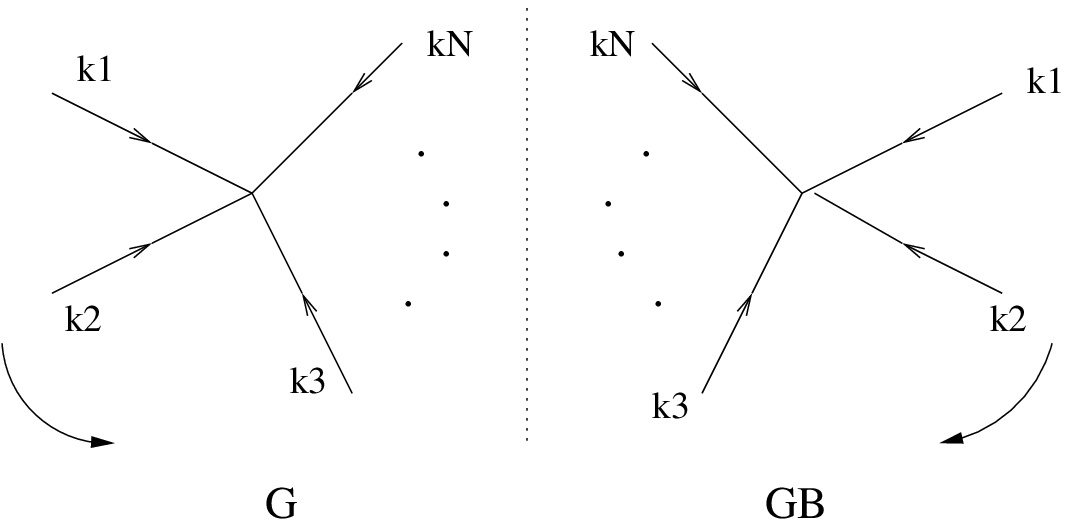}
\caption{A Feynman diagram $G$ and its mirror diagram $\bar{G}$. They have
the opposite Moyal phase factors.}
}

In a noncommutative theory, the Feynman amplitude for a diagram 
$G$ takes the form
\be
M_{ab}^G(s,\st) = \prod_{l,i} \int \frac{d^D k_l \; B}{D_i^+} V^G.
\ee
Here $1/D_i^{\pm}$ is the propagator of the $i$-th internal line and the
mass square
has a small $\mp$ imaginary part and $B$ is  a real normalization
factor that contains the couplings
\footnote{
We emphasis that the couplings (bare as well as the renormalized
one) have to be real. As we  discuss at the end of section 3.1,
the original Lee model (defined in 
4-dimensional spacetime and with the dispersion relations
\eq{luchu1.9}) has an imaginary bare coupling \cite{lee} and  Hermitian
analyticity does not hold, in both the commutative and noncommutative cases. 
However, the improved Lee models have real couplings and so have hermitian
analytic  $S$-matrix.
} 
and factors of $\pi$, $i$ etc. 
Complex conjugating, one has
\be
(M_{ab}^G(s,\st))^* = \prod_{l,i} \int \frac{d^D k_l \; B}{D_i^-} (V^{G})^* 
= M_{ba}^{\bar{G}}(s^*,\st^*).
\ee
where we have used in the last step the
observation that a change of sign in the imaginary part
of the mass (or cutoff) corresponds to the
 change of sign in the imaginary part of
$s$ (or $\st$). 
In the discussion given above, for the clarity of the argument, 
we have been careful to indicate which diagram ($G$ or
$\bar{G}$) is to be drawn for the Feynman amplitude to be computed.
However this is not really necessary as which diagram has to be
drawn is already clear once the the process to be considered 
($a \to b$ or $b \to a$) is specified. Therefore, can simply write
\be \label{ncHA}
(M_{ab}(s,\st))^* = M_{ba}(s^*,\st^*).
\ee
Thus we have shown that the Feynman diagrams (and hence the
$S$-matrix) of a noncommutative  theory are
hermitian analytic. 
We stress that our result is general and does not depend on the
detailed form of the propagators or vertices. For example, it applies
to the noncommutative Lee model to be introduced in section 3.

\subsection{Unitarity Constraints and their Relation to the IR/UV Singularities}

Note that the symmetric condition \eq{symmetric} is, in
general, not valid and so the condition  \eq{img} may not hold even if a theory is unitary. 
However, since Feynman amplitudes satisfy hermitian analyticity, 
\eq{discT} and
\eq{discM} hold if the $S$-matrix is unitary. 
Therefore we
propose to use \eq{discT} or \eq{discM} instead of \eq{img} 
\footnote{
In \cite{gomis}, the $1\to 1$ propagator diagram in the
noncommutative $\phi^3$ and the $2 \to 2$ scattering diagram in the
noncommutative $\phi^4$ were considered. It is easy to see that the
symmetric condition \eq{symmetric} is satisfied for these processes, and
so checking of \eq{img} constitutes a valid test of unitarity for
the noncommutative theories considered there. }
as a check of unitarity.

Before we consider a specific model, let us
discuss how the IR/UV singularities may lead to a breakdown of
unitarity in general. Generally,
a new IR/UV singularity in a scattering amplitude can be a pole or 
a branch point in $\st =0,$ for some $\st$. 
Note that 
\be
\st = (\th_E)^2 (p_0^2 -p_1^2)+ (\th_B)^2 (p_2^2 +p_3^2),
\ee
where we have chosen, for example, $\th^{01} = \th_E$, $\th^{23}=
\th_B$ with all other components vanishing.
Therefore for space noncommutativity, $\st$ is positive definite and
so there is no new contribution to the discontinuity of the amplitude
from this singularity. However, in the case of space-time noncommutativity,
$\st$ is not of definite sign in the physical region \cite{gomis}. 
Therefore if $\st$ is a branch point singularity,
there will now be a new contribution to the LHS of
\eq{discM}. 
Since the IR/UV singularities do not correspond to any
intermediate degrees of freedom that can go on shell, 
these new contributions will not be accounted for by the ``onshell'' sum
and \eq{discM} will be violated. 
This is the basic mechanism how unitarity is violated 
by the IR/UV singularities  when time is noncommuting.
Both the IR/UV singularity and the noncommuting time 
must be present in order to violate unitarity.  
Finally, we would like to add, as was shown in \cite{agbz}, that
even when one tries to add new degrees of freedom
to satisfy the cutting rules in a formal sense,
these new degrees of freedom have to be tachyonic and so the theory is 
inconsistent.

\section{An Application: The Noncommutative Lee Model}

In this  section, we  consider the Lee model in $D$ spacetime
dimensions and its noncommutative generalization. In particular, we 
consider the issues of the IR/UV mixing and unitarity
for the noncommutative Lee
model. We will find that due to the presence of the  Moyal phase
factors the symmetric condition is not satisfied. 
Therefore one should check unitarity using \eq{discM}.  
We  show that (and this result is exact)  
the noncommutative Lee model is free from any  IR/UV singularity.
As a result, one can expect that the noncommutative Lee model  is unitary for
both the space-space and space-time noncommutative case. We give
further arguments supporting this claim.

Another model which is free from the IR/UV mixing is the noncommutative Chern-Simon 
model. This model is finite and, as shown by \cite{bgps},
 free from the IR/UV mixing at the one 
loop level. However, it is actually a free theory, 
at least in the axial gauge \cite{ds}. Thus this
model is not suitable for  our purposes.

\subsection{Commutative Case}
\setcounter{equation}{0}

The Lee model was originally introduced by Lee in \cite{lee} 
where it was shown that the model is renormalizable with its mass,
wavefunction and charge renormalizations easily performed in an
exact manner. 
In the following, we follow the presentation of \cite{schweber}.
The model has 
two fermions $V$ and $N$
with masses  $m_{V}^{(0)}, m_{N}^{(0)}$ respectively,
and a real scalar $\vphi$ with mass $m_{\vphi}^{(0)}:= \mu_0 $.
The Hamiltonian for the free fields is:
  \be \label{luchu2.3}
  H_0 =  \int d^{D-1} \, {\bf p}\, \left[
E_V ( {\bf p})  \, V\dag ( {\bf p}) V( {\bf p}) +
E_N ( {\bf p})  N\dag ( {\bf p} ) N ( {\bf p} ) + 
E_\vphi ( {\bf p}) \vphi\dag ( {\bf p} ) \vphi ( {\bf p}) \right],
\ee
where $E_V ( \bf{p})$, $E_N( \bf{p})$, $E_\vphi ( \bf{p})$ are the
dispersion relations for the  free  $V,N$ and $\vphi$ particles, 
$N(\bp), V(\bp)$ and $\vphi(\bp)$ are 
the annihilation operators of the $N, V$ and $\vphi$ particles, respectively.
In the original Lee model \cite{lee}, $D=4$ and 
the fermions are taken to be very heavy while $\vphi$  is
assumed to be relativistic. In this case, the  dispersion relations are given by
\be \label{luchu1.9}
E_V = m^{(0)}_{V} \, ,
\qquad
E_N = m^{(0)}_V \, , \qquad
E_\vphi ( {\bf k}) = ( {\bf k}^2 + \mu_0^2  )^{1/2} := \omega_{\bf k}.
\ee 
The Galilei-invariant form \cite{lev16}
\be \label{luchu2.4a}   
E_A({\bf p}) = \frac{{ \bf p}^2}{2m^{(0)}_A}, \qquad A=V,N,\vphi,
\ee
as well as the relativistic choice \cite{ydu15}
\be \label{luchu2.4b}
   E_A ( {\bf p} ) = \left( {\bf p}^2  + {m^{(0) \; 2}_A} \right)^{1/2}
\ee
were also studied in the literature.
The interacting Hamiltonian of the model is taken to be given by 
\be \label{luchu1.10}
H_{\rm int} = g_0\int \frac{d^{D-1} \bk}{\sqrt{(2 \pi)^{D-1}\; 
2 \o_{\bk}} }\int 
d^{D-1} \bp  
\left( V^\dag(\bp) N (\bp-\bk) \vphi(\bk) \; f({\bf k})  
+ \vphi\dag(\bk) N\dag (\bp-\bk) V(\bp)  \; f^*({\bf k}) 
\right),
\ee
where $f(\bk)$ is a form factor
\footnote{ 
Note that, in principle, one can also use a more general 
form factor $f$ that depends  on the momentum of the $N\vphi$ pair.
It is easy to see that this amounts to a simple replacement 
\be \label{luchu2.5}
     f( {\bf k}) \longrightarrow f (k,p).
\ee
in the  analysis below.} 
introduced to smooth out the
interaction to avoid the divergences connected with a
point interaction. In fact $f$ can be taken to be $f=1$ and the divergences can
be absorbed by renormalization. This is the case of interest to us.
However as we will see,
the introduction of noncommutativity to the Lee model amounts to a
modification  of $f$ by a phase factor. 
Therefore we will  keep $f$ explicitly in the presentation below, with
the understanding that it will be set to 1 (or to the Moyal phase
factor for the noncommutative case) in the final answer.

We note  that the interaction $H_{\rm int}$
is nonlocal in space even in the limit $f=1$. To see this,  it is convenient to
introduce the negative and positive frequency parts of $\vphi$:
\be
\vphi(x)= a(x)+ a\dag(x),
\ee
\be
a(x) = \int \frac{d^{D-1} \bk}{\sqrt{(2\pi)^{D-1}\; 2 \o_\bk}} \vphi(\bk) 
e^{i k \cdot x}, \quad 
a\dag(x) = \int \frac{d^{D-1} \bk}{\sqrt{(2\pi)^{D-1}\; 2 \o_\bk}}
\vphi\dag(\bk) 
e^{-i k \cdot x}.
\ee
In terms of $a$ and $a\dag$,  $H_{\rm int}$ can be written in the
coordinate space as
\be
H_{\rm int} = g_0\int d^{D-1} \bx \; d^{D-1} \by 
\left( V\dag(\bx,t) N(\bx,t) \ft(\bx-\by) a(\by,t) + N\dag(\bx,t)
V(\bx,t) \ft^*(\bx-\by) a\dag(\by,t) \right),
\ee
where $ \ft$ is the Fourier transform of the Lee model
form factor and $\ft \to \delta(\bx)$ in the limit $f \to 1$.  
It is now clear that the coupling term is nonlocal in space 
since the operation of taking the positive frequency part 
involves the integration over all space. 
However the model is local in time.

Since the theory is local in time, it can be 
described equivalently in the Lagrangian formulation by performing
the Legendre transformation. 
The Lagrangian density of the model is given by
\be
\label{luchu1.6}
{\cal L} = {\cal L}_0 + {\cal L}_{\rm int},
\ee
where ${\cal L}_0$ is the free part:
\be
 \label{luchu1.7}
 {\cal L}_0  =  V\dag 
\left( i \frac{\partial}{\partial t} + E_V(-i{\bf \nabla}) \right) V
+ N\dag 
\left( i \frac{\partial}{\partial t} + E_N( -i  {\bf \nabla}) \right) N
+ a\dag 
\left( i \frac{\partial}{\partial t} + E_\vphi(  -i {\bf \nabla} ) \right) a,
\ee
and the interaction is described by
\be  \label{luchu1.8}
{\cal L}_{\rm int} =
g_0\int \, d^{D-1} {\bf y} \;V\dag ( {\bf x}, t )
N ( {\bf x}, t )
\widetilde{f} ( {\bf x} - {\bf y} )
a ( {\bf y} , t) + H.C.
\ee
The Lagrangian formulation 
will be useful when we introduce an electric deformation of the model.

The Lee model can be solved by considering 
directly the Schr\"{o}dinger equation with the
Hamiltonian $H = H_0 + H_{\rm int}$ 
where $H_0$ is given by (\ref{luchu2.3}) 
with the choice (\ref{luchu1.9}) 
and $H_{\rm int}$ is given by (\ref{luchu1.10}).
Due to the structure of the interaction \eq{luchu1.8},
the only elementary interaction of the theory involves the process
\be \label{nit}
V \stackrel{\rightharpoonup}{\leftharpoondown} N+ \vphi.
\ee
In a standard relativistic model, the antiparticle $\bar{\vphi}$ would
appear and the crossed reaction  
\be \label{luchu2.6}
V + \bar{\vphi} \stackrel{\rightharpoonup}{\leftharpoondown} N
\ee
would be possible, but  this is not allowed in the Lee model due to
the particular form of the interaction Hamiltonian \eq{luchu1.10}.
The system possesses two simple conservation laws
\be \label{conser}
n_V + n_N =\mbox{constant}, \quad
n_V + n_\vphi = \mbox{constant},
\ee
where $n_V, n_N, n_\vphi$ are the total numbers of $V,N,\vphi$ particles, 
respectively.
Due to the conservation laws \eq{conser}, the eigenfunctions of $H$
contain only a finite number of particles and,  consequently,
the theory is exactly solvable \cite{lee}. 

\vs
\underline{Renormalization}
\vs

The quantization of the theory is straightforward.
Locality in time allows us to perform the standard  canonical quantization 
of the theory.
The nontrivial commutation relations of the field operators are
\be
[\vphi(\bk), \vphi\dag({\bk'})] = \d (\bk -\bk'), \quad
[N(\bp), N\dag(\bp')]_+ = \d(\bp -\bp'), \quad
[V(\bp), V\dag(\bp')]_+ = \d (\bp -\bp'),
\ee
with the rest equal to zero.
The vacuum of the theory $\ket{0}$ is defined by
\be
N(\bp)\ket{0} = V(\bp)\ket{0} =\vphi(\bp) \ket{0} =0.
\ee
It is easy to verify that
\be \label{luchu2.10}
H_{\rm int} \vphi\dag(\bk) \ket{0} =0, \quad H_{\rm int} N\dag(\bp) \ket{0} =0;
\ee
thus we can take the $\vphi$ and $N$--quanta as the physical particles
(of masses $\m$ and $m_N$, respectively) and identify $\m= \m_0$, $m_{N_0} = m_N$, and
there is only the renormalization of the mass of $V$ to be considered.

Without any loss of generality we consider the dispersion relation s
\eq{luchu1.9} in order to study the renormalization of the theory.
Consider
the sector of the theory associated with one physical $V$-particle. Denote the
physical V-particle as $\ket{\Vh(\bp)}$.
Due to the conservation law \eq{conser}, we have
\be \label{physicalV}
\ket{\Vh(\bp)} = \sqrt{Z_V} \left(V(\bp)\dag \ket{0}+ \int
d^{D-1}\bk \; \Phi(\bk) \; N\dag(\bp -\bk) \vphi\dag(\bk)   \ket{0} \right)
\ee
with the wavefunction $\Phi(\bk)$ still to be determined.  
Here $\ket{\Vh(\bp)}$ is an eigenstate of $H$
\be \label{HV} 
H \ket{\Vh(\bp)} = m_V \ket{\Vh(\bp)}.
\ee
The normalization of $\ket{\Vh(\bp)}$ yields
\be \label{p1}
1= Z_V (1+ \int d^{D-1} \bk \; |\Phi(\bk)|^2 ).
\ee
Contracting \eq{HV}  with $\bra{0} V(\bp')$, one obtains
\be \label{p2}
m_{V_0} + \frac{g_0}{(2 \pi)^{(D-1)/2}}\int \frac{d^{D-1} \bk}{\sqrt{2 \o_\bk}}
f (\bk) \Phi(\bk) = m_V.
\ee
On the other hand, contracting \eq{HV} with $\bra{0}N(\bq) \vphi(\bl)$, one obtains
\be \label{luchu2.15}
(m_V- m_N -\o_\bk) \Phi(\bk) = \frac{g_0}{(2 \pi)^{(D-1)/2}}
\frac{f^*(\bk)}{\sqrt{2 \o_\bk}},
\ee
which gives
\bea
\Phi(\bk) &= \frac{g_0}{(2 \pi)^{(D-1)/2}}
\frac{f^*(\bk)}{\sqrt{2 \o_\bk} (m_V- m_N -\o_\bk)}, \quad
& \mbox{for $m_V< m_N +\mu$}, \label{p3} \\
\Phi(\bk) &= \frac{g_0}{(2 \pi)^{(D-1)/2}} \mbox{P}
\frac{f^*(\bk)}{\sqrt{2 \o_\bk} (m_V- m_N -\o_\bk)}, \quad
& \mbox{for $m_V>m_N +\mu$}. \label{p4}
\eea
Note that eq. \eq{p3} corresponds to the case when the 
$V$ particle is stable; i.e. it cannot
spontaneously decay  into an $N$ and $\vphi$ particle.
The  decay of the $V$ particle is allowed in the case of eq. \eq{p4}. 
The renormalized coupling can be obtained by requiring the scattering process
\be
N + \vphi \to N + \vphi
\ee
to be nonzero in the limit  $f\to 1$.

As a result, we obtain the following renormalization constants
\bea 
Z_V^{-1} &=& 1 + \frac{g_0^2}{(2 \pi )^{D-1} }
\int  \frac{d^{D-1}\bk}{2\omega_\bk}
\,
\frac{|f(\bk)|^2}{(m_V - m_N -\omega_\bk)^2},
\label{luchu2.18} \\
m_V &=& m_{V_0} + \frac{g_0^2}{(2 \pi )^{D-1} }
\int  \frac{d^{D-1} \bk}{2\omega_\bk}
\,
\frac{|f(\bk)|^2}{(m_V - m_N -\omega_\bk)},
\label{luchu2.19}\\
g^2 &=& g_0^2 Z_V. \label{luchu2.20}
\eea 
The integrals in \eq{luchu2.18} and \eq{luchu2.19} are generally
divergent in the limit $f\to 1$. As usual, all the scattering amplitudes 
($N\vphi-N\vphi$, $V\vphi-V\vphi$,   
$V\vphi-N\vphi\vphi$, $N\vphi\vphi-N\vphi\vphi$ etc.) become finite
after we have  performed the renormalization \eq{luchu2.18}, \eq{luchu2.19}
and \eq{luchu2.20}\footnote{For example, in the sector $V\vphi-N\vphi\vphi$,
the renormalized scattering amplitudes $V\vphi \rightarrow V\vphi$,
$V\vphi\rightarrow N\vphi\vphi$ and 
$N\vphi\vphi\rightarrow N\vphi\vphi$ were studied in \cite{amad} and \cite{fuda}}.

\FIGURE{
\label{fey}
\psfrag{A}{$A$}
\psfrag{Ad}{$A\dag$}
\psfrag{p}{$p$}
\psfrag{k}{$k$}
\psfrag{f1}{$\frac{1}{p_0 -E_A(p_0,\bp) + i \e},\;\; A=V,N,\vphi$}
\psfrag{V}{$V$}
\psfrag{N}{$N$}
\psfrag{phi}{$\vphi$}
\psfrag{Vd}{$V\dag$}
\psfrag{Nd}{$N\dag$}
\psfrag{phid}{$\vphi\dag$}
\psfrag{f2}{$g_0 f(k,p)$}
\psfrag{f3}{$g_0 f^*(k,p)$}
\psfrag{f4}{for each loop momentum integration: 
\hspace{0.3cm}  $(2\pi)^{-D} \int d^D k$}
\includegraphics{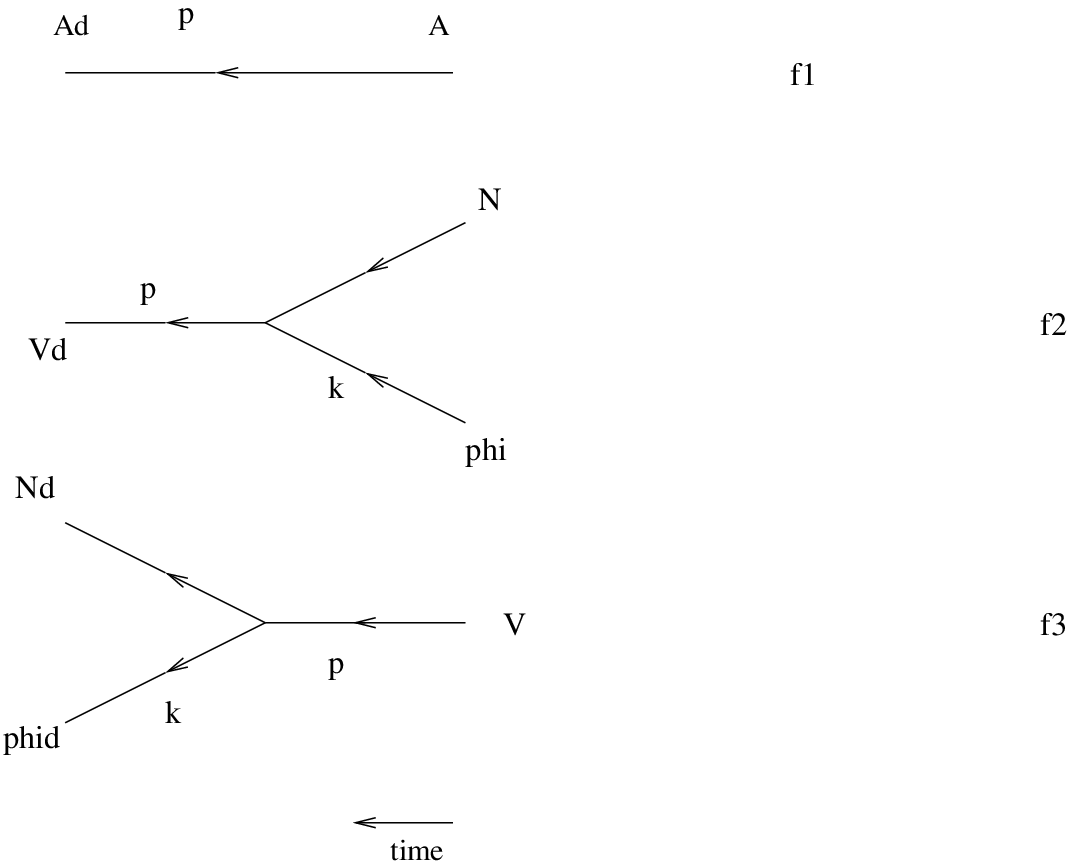}
\caption{Feynman rules for the Lee model}
}

We would like to add a couple of comments:

i) One can perform a 
path integral quantization of the theory and one obtains the 
Feynman rules given in figure 2. Using these Feynman rules, 
it is straightforward to show that the above results for the renormalization  can  
also be obtained in the Lagrangian framework and are exact in perturbation theory.
Later we will use these Feynman rules to study the
noncommutative  Lee model, particularly, in the time noncommuting case.

ii) In the original Lee model \cite{lee}, $D=4$ and 
the mass renormalization constant is  
linearly divergent while the wavefunction renormalization is 
logarithmically divergent. 
It has been shown that the different
choices \eq{luchu2.4a} (Galilean kinematics)
and \eq{luchu2.4b} (relativistic kinematics) 
of dispersion relations  lead
to finite renormalizations when $ f \to 1 $.

\vs
\underline{Unitarity and the ghost state}
\vs

The relation \eq{luchu2.20} 
between the renormalized coupling $g$ and the bare coupling $g_0$ can
be rewritten as 
(with $f$ set to 1)
\be \label{gI}
g_0^2 = \frac{g^2}{1- g^2 I}, \quad {\rm where} \quad
I \equiv
\int  \frac{d^{D-1}\bk}{(2 \pi )^{D-1}\; 2\omega_\bk}
\,
\frac{1}{(m_V - m_N -\omega_\bk)^2} >0.
\ee
For $D=4$, $I$ is logarithmically divergent. If $g$ is to remain fixed 
and nonvanishing,  the bare coupling has to be imaginary
\be \label{g1}
g_0 = i \infty^{-1}.
\ee
and the wavefunction renormalization, 
\be \label{Z1}
Z_V = 1 - g^2 I \quad \to - \infty.
\ee
This contradicts the interpretation of $Z_V$  as the probability of
finding a bare $V$ quantum in the physical $V$-particle state.  
Such negative probabilities imply that the $S$-matrix is not unitary.
In fact one can show that \cite{pauli} $Z_V <0$ corresponds to a new
state in the theory. This state $\ket{G}$ has a negative norm  and is
referred to as the ``ghost state'' by Kallen and Pauli. As a result, the
$S$-matrix is explicitly non-unitary.  
In fact, the not unitarity of the theory is related to  the 
original Hamiltonian being non-Hermitian due to the
presence of an imaginary bare coupling. 

Two improvements of the original Lee model are possible. One is to
consider other dispersion relations  e.g. \eq{luchu2.4a} and
\eq{luchu2.4b}.  This leads
to 4-dimensional theory with finite renormalizations and without
a ghost \cite{lev16,ydu15}. Another possibility is to consider the Lee model
in lower dimensions \cite{bn}. In $D=3$, the integral $I$ in \eq{gI} is 
finite and so the model is ghost free for physical coupling 
$0<g< 1/\sqrt{I}$. 
The  improved Lee model is still exactly solvable in both cases.
To minimize the number of new formulae, we 
consider the second class of models 
when we generalize to the noncommutative case.

\subsection{The Noncommutative Lee Model}

The noncommutative framework is  generated by using the $*$-product
\eq{luchu1.5}. As mentioned in the introduction,  the noncommutative
deformation can be introduced either in the Hamiltonian or the
Lagrangian formulation in the magnetic case ($\theta^{i0} = 0$).  
The replacement (\ref{luchu1.5}) 
amounts to the following substitution in the formula  (\ref{luchu1.10}):
\be  \label{luchu1.11}
 f( {\bf k}) \longrightarrow 
f(k,p ) := f( {\bf k} ) \,
 e^{ \frac{i}{2} p_i \theta^{ij} k_j }.
\ee
In  the electric case with nonvanishing
components $\theta^{i0} \neq 0$ 
\footnote{
Besides magnetic and electric cases
one can also consider lightlike deformations \cite{lightlike}, 
corresponding to the  case $\theta_{\mu\nu}\theta^{\mu\nu}=0$.
},
the substitution  takes the form
\be \label{luchu4.1}
f (\bk ) \longrightarrow f(k,p):=f (\bk) \, 
e^{ \frac{i}{2} \theta^{0i} ( p_0 k_i -p_i \, k_0 )}.
\ee
Obviously the
$*$-product involves an  infinite number of time derivatives. 
The nonlocalities in time destroy not just
the usefulness of the Hamiltonian formulation, but also 
the standard way of relating the Lagrangian and the Hamiltonian description
\footnote{For recent efforts at introducing  a Hamiltonian framework for Lagrangian
 densities nonlocal in time see \cite{llo19,woo20,gon21}. 
We have not been able to employ these results here in a constructive way.}.
We are thus left only with the Lagrangian framework. 
For example, when there is only the nonvanishing component 
$\theta^{01} = \th \neq 0$, one obtains the modification  of the
product of $V$ and $\vphi$ fields 
\be
\label{luchu4.2}
V( {\bf x}, t) * a ( {\bf y}, t ) =
 e^{-i \frac{\theta}{2}( \frac{\partial}{\partial t}\,
 \frac{\partial}{\partial y_1}
 - \frac{\partial}{\partial x_1}\,
 \frac{\partial}{\partial t'} )
 }
V( {\bf x}, t ) a ( {\bf y}, t' )|_{t=t'} =
V    ( {\bf x}, t - \frac{i\th}{2} \frac{\partial}{\partial y_1} )
a ( {\bf y}, t + \frac{i\th}{2}\overleftarrow{\frac{\partial}{ \partial x_1}}
)
\ee
in the interaction Lagrangian \eq{luchu1.8}.
Note that 
due to the associativity of the Lagrangian and the integration over spacetime,
the $*$-product of the three fields in \eq{luchu1.8} can be represented by a
modification of the product for any pair of fields ($V\vphi$ as in \eq{luchu4.2},
$VN$ or $N\vphi$). 

Note also that the phase factor in \eq{luchu1.11} and \eq{luchu4.1} 
does not lead to a real factor as in the noncommutative
scalar $\phi^3$ case.
Thus the noncommutative modification in the Lee model involves a complex 
factor. This, in particular, implies that the symmetric condition
\eq{symmetric} is not satisfied.

Quantization of the magnetically deformed theory can be achieved by
using either the canonical quantization, or equivalently  a path integral
quantization.  In the electric case,
canonical quantization fails due to the nonlocality in time. 
Nevertheless, formally, the theory can be quantized  using the path integral
method. In the following, we will use the path integral method to
analyze both the magnetic and the electric Lee models. 
The Feynman rules are those of figure 1 with
$f(k,p)$ given by \eq{luchu1.11} and \eq{luchu4.1} and work for
general $D$. To be specific, below we  consider the noncommutative
Lee model in $D \leq 4$ dimensional spacetime and  
with the standard dispersion relations \eq{luchu1.9}.

\vs 
\underline{Renormalization and (no) IR/UV mixing} 
\vs

Since the effect of noncommutativity is a modification
\eq{luchu1.11}  or \eq{luchu4.1}  of $f$ by a phase
factor, it is clear that the mass, wavefunction and  coupling renormalization
(depending on $|f|^2$)
are not affected. Thus we conclude that the renormalization constants
of the noncommutative Lee model are
exactly computable and are independent of the noncommutativity
parameter $\th$.

Moreover, one can easily convince onself that the UV-divergences 
of the theory reside in planar diagrams that simply do not have 
nonplanar counterparts.
Thus the UV-divergences of the noncommutative Lee model remain
untouched in the limit when the cutoff is removed. 
This is quite different from the other noncommutative field theories
which display an intriguing mixing of IR/UV \cite{uvir}. In these
models, the introduction of a nonzero noncommutativity improves the UV
convergence of  nonplanar diagrams but also leads to new IR
singularities for these diagrams.  
In the present case of the noncommutative Lee model, there  simply are
no UV-divergences in the nonplanar diagrams, and hence there 
are also  no
new IR singularities that could be generated. We conclude that 
the noncommutative Lee model is {\it free from IR/UV mixing}. This result is exact.

\vs
\underline{Unitarity}
\vs

First we consider the unitarity constraints at the one loop level. 
Due to the structure of the vertices (figure 2) in the theory, it is easy to convince
oneself that only planar diagrams can be drawn at the one loop
level. Therefore
the one loop Feynman amplitudes take the form
\be \label{mm}
M_{ab} = M_{ab}^{(0)} e^{i \phi_{ab}},
\ee
where $M_{ab}^{(0)}$ are the corresponding amplitudes in the commutative
case, and $e^{i\phi_{ab}}$ is the Moyal phase factor associated with the
planar diagram. As a result, the equation \eq{discM} is satisfied since
\be
{\rm Disc} M_{ab} = e^{i \phi_{ab}} {\rm Disc} M_{ab}^{(0)} =  
i e^{i \phi_{ab}} \sum_n M_{an}^{(0)} M_{nb}^{(0)} = 
i \sum_n M_{an} M_{nb}.
\ee
In the second step, we have used the fact  that the constraint
\eq{discM} is satisfied for the commutative Lee model since this model is
unitary (or one can verify this in a straightforward manner since
the $M$'s that appear in the sum are tree level ones). In the
last step we have used the fact that the planar Moyal phase factor of
the 1-loop diagram decomposes simply into the product of factors of 
the tree level ones:
\be
e^{i \phi_{ab}} = e^{i \phi_{an}}e^{i \phi_{nb}}.
\ee
Note that due to the form of the modification for the one-loop amplitude \eq{mm}, 
checking the imaginary part \eq{img} would lead to
the incorrect conclusion that the noncommutative Lee model is not
unitary at a one loop level.
Note also that the above argument is general and does not depends on 
whether $\th$ is spacelike or timelike. 
Therefore, we conclude that the noncommutative Lee model is unitary at
a one loop level for general $\th^{\m\n}$. This result is valid to all
orders in $\theta$.

\FIGURE{
\psfrag{p1}{$p_1$}
\psfrag{p2}{$p_2$}
\psfrag{p3}{$p_3$}
\psfrag{p4}{$p_4$}
\psfrag{k}{$k$}
\psfrag{V}{$V$}
\psfrag{Vd}{$V\dag$}
\psfrag{t}{$t$}
\includegraphics{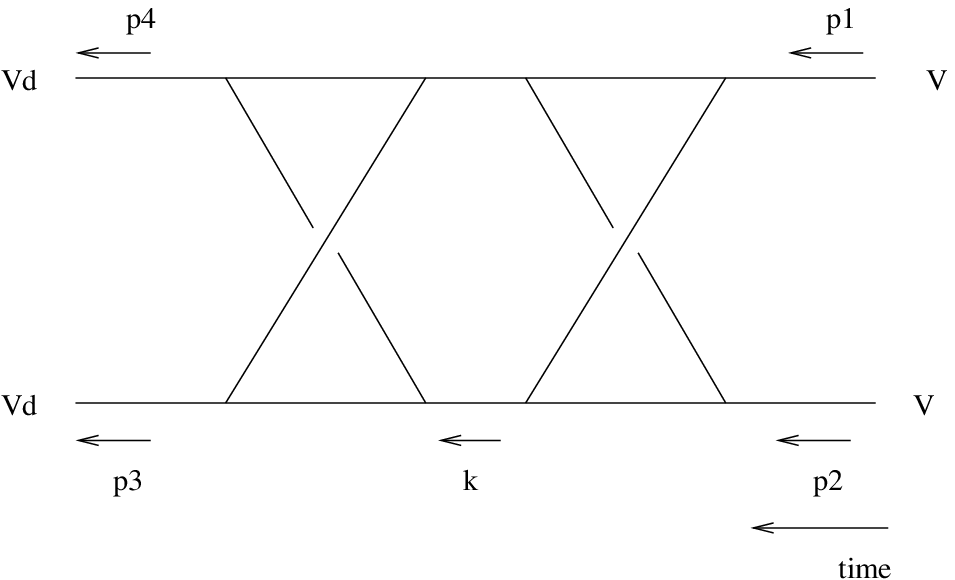}
\caption{A nonplanar diagram}
}

At a higher loop level, one can have nonplanar diagrams, for example,
the one in figure 3. 
The phase factor associated with this diagram is
\be \label{mpp}
e^{\frac{i}{2} (p_1 \th p_2 - p_1 \th p_3 - p_2 \th p_3)}  e^{-i k \th (p_2 -p_3)}.
\ee
The second phase factor depends on the loop momentum and is a characterization
of a nonplanar diagram. As one can check easily,
this  amplitude is regular in the variable $\st$ (and hence
$\th$). Generally, due to  the absence of the IR/UV singularity,
a nonplanar amplitude will be regular in the variable $\st$ and so
there is no new discontinuity in the LHS of the unitarity equation
\eq{discM}.
Since both the LHS and RHS are regular in $\th$, 
the unitarity constraint will be satisfied at the zeroth order
in $\theta$. 
Although we believe this to be the case, 
it may not be easy to verify  the  unitarity relations
to all orders in $\th$ as one would have to exploit various
nontrivial relations among special functions and integrals.
The fact that  unitarity constraints 
are satisfied at a one loop level; and also (at the zeroth order in
$\th$) for any  higher loop amplitude, is already a nontrivial property of the
noncommutative Lee model. 
Without any other source of violation of unitarity in sight, we 
expect that the noncommutative Lee model is
unitary for any $\th^{\mu \nu}$.

\section{Discussion}

In this paper, we have discussed and examined two basic aspects of 
noncommutative field
theories: the IR/UV mixing and unitarity. 
We have showed that the $S$-matrix of a noncommutative
field theory is hermitian analytic. This implies that unitarity
provides a direct evaluation of the discontinuities associated with the
cuts of  normal thresholds. We  have also
 explained how the IR/UV singularities can lead to a
violation of unitarity for field theories with  space-time
noncommutativities. As a corollary, we have argued that a
noncommutative field theory without any IR/UV mixing will
be unitary in both the space-space and space-time
noncommutative cases.

As an illustration of the  general discussion, 
we have introduced and analysed the noncommutative Lee model.
We have found that the model is entirely free from the IR/UV mixing 
This result is exact. Our general arguments show that  the
noncommutative Lee model is unitary in both the space-space and space-time
noncommutative cases. Simple explicit checks are  consistent with this claim.
Thus we provide a counter example to the general
belief that field theories with space-time noncommutativity have to be
non-unitary.

A consistent quantum field theory on a noncommutative spacetime should
be unitary. It should also be free from the problems related to the IR/UV
mixing. 
One can broadly divide the IR/UV  mixing phenomena in noncommutative field
theories into those that could be called good ones and  bad ones. For
example, the IR/UV singularities which appear in a purely bosonic
noncommutative gauge theory or in a noncommutative QED are bad ones \cite{susskind}.
However, IR/UV singularities are milder and  may
be absent \cite{k1} in the presence of  supersymmetry.
The milder form of the IR/UV mixing in supersymmetric noncommutative gauge
theories leads to a decoupling of the $U(1)$
degrees of freedom in the IR \cite{AMT,HKT}.
Not only the $U(1)$ degrees of freedom become free in the IR \cite{HKT},
they also trigger spontaneous supersymmetry breaking \cite{ckt}
in the presence of an appropriate Fayet-Iliopoulos D-term
and play the r{\^o}le of the hidden sector. This we refer to as  good IR/UV
mixing effects. More details are provided  in \cite{ckt2}. 
With unitarity better understood and (some) IR/UV mixing turned to be our advantage,
it seems not unreasonable to contemplate that nature could indeed be 
noncommutative (at least at some level of explanation of its phenomena).

\acknowledgments{CSC and JL would like to thank Luis Alvarez-Gaume for useful
discussions and comments and 
the theory group at CERN, where this work was started, for its hospitality.
We would also like to thank David Fairlie, 
Pei-Ming Ho, Valya Khoze, Rodolfo Russo, Lenny Susskind, 
Richard Szabo and Gabriele Travaglini for helpful discussions.
}

\ed